\documentclass[runningheads]{llncs}
\usepackage[T1]{fontenc}
\usepackage{graphicx,verbatim}
\usepackage{amsmath}
\usepackage{graphicx} 
\usepackage{caption} 
\usepackage{subcaption} 
\usepackage{float} 
\usepackage{epstopdf}   
\usepackage{amssymb}

\usepackage{subfig}     
\usepackage{array}      

\begin{document}
%
\title{Diffusion-based Virtual Staining from Polarimetric Mueller Matrix Imaging}
\author{Xiaoyu ZHENG\inst{1} \and
Jing WEN\inst{5} \and
Jiaxin ZHUANG\inst{1} \and
Yao DU\inst{1} \and
Jing CONG\inst{2} \and 
Limei GUO\inst{3} \and
Chao HE\inst{4} \and
Lin LUO\inst{5} \and
Hao CHEN\inst{1} }
\authorrunning{Xiaoyu Zheng et al.}
%
\institute{The Hong Kong University of Science and Technology, Hong Kong SAR, China \and
Beijing Institute of Collaborative Innovation, China \and
Peking University Health Science Center, China \and
University of Oxford, United Kingdom \and
Peking University, China
}

%
%



\maketitle              
\begin{abstract}
Polarization, as a new optical imaging tool, has been explored to assist in the diagnosis of pathology. Moreover, converting the polarimetric Mueller Matrix (MM) to standardized stained images becomes a promising approach to help pathologists interpret the results. However, existing methods for polarization-based virtual staining are still in the early stage, and the diffusion-based model, which has shown great potential in enhancing the fidelity of the generated images, has not been studied yet. In this paper, a Regulated Bridge Diffusion Model (RBDM) for polarization-based virtual staining is proposed. RBDM utilizes the bidirectional bridge diffusion process to learn the mapping from polarization images to other modalities such as H\&E and fluorescence. And to demonstrate the effectiveness of our model, we conduct the experiment on our manually collected dataset, which consists of 18,000 paired polarization, fluorescence and H\&E images, due to the unavailability of the public dataset. The experiment results show that our model greatly outperforms other benchmark methods. Our dataset and code will be released upon acceptance.


\end{abstract}

\section{Introduction}
Polarization imaging technology enables non-invasive acquisition of birefringence characteristics and surface microstructure information at sub-resolution scales of tissue structures by analyzing the modulation properties of biological tissues on the polarization state of incident light waves (such as Stokes vectors, Mueller matrices, and other parameters)~\cite{ref1}. This unique imaging mechanism imparts high sensitivity in detecting pathological features such as collagen fiber alignment and cell membrane integrity, demonstrating significant potential for clinical diagnosis applications in recent years, including tumor boundary delineation and tissue fibrosis evaluation\cite{ref2,ref3}. However, due to the human visual system's lack of direct perception of polarization states, the rich physical information contained in polarization images needs to be converted into grayscale or pseudo-color images through complex post-processing algorithms for visualization. This secondary encoding process results in significant differences between the imaging results and the cognitive paradigms of pathologists, severely limiting the interpretability and acceptability of polarization imaging technology in clinical diagnosis~\cite{ref4}.

In traditional pathological practice, fluorescence imaging and Hematoxylin and Eosin (H\&E) staining techniques have become the gold standard due to their clear biological relevance. Fluorescence labeling enables functional imaging at the molecular level through specific antibody-antigen reactions~\cite{ref5}, while H\&E staining provides intuitive histomorphological information through differential staining of cell nuclei and cytoplasm~\cite{ref6}. 
The emergence of virtual staining technology has offered novel solutions to improve the efficiency of computational pathology. This technology establishes a non-linear mapping relationship between unstained tissue images and target staining modalities through deep learning models, achieving significant advances in the field of computational pathology in recent years~\cite{ref7}. Mainstream methods primarily rely on Generative Adversarial Networks (GANs) and their variants, realizing cross-modal image translation via Cycle-Consistency constraints~\cite{ref8}. Recent research has further explored the application of Diffusion Models in virtual staining, leveraging their progressive denoising mechanism to enhance the morphological fidelity of staining boundaries~\cite{ref9}. However, diffusion methods for polarization data has not been investigated in previous works. 
In this study, we integrate the Mueller matrix full-polarization imaging system with a diffusion-based virtual staining framework to establish an image-to-image translation model from raw polarization features to standard Hematoxylin and Eosin (H\&E) as well as fluorescence images, namely the Regulated Bridge Diffusion Model (RBDM). The contributions are summarized in the following aspects:

\textbf{(1) First Public Multi-modality Polarization Pathology Dataset}: We construct and release a benchmark dataset that contains 18,000 strictly registered image patches with polarization, H\&E, and fluorescence modalities.


\textbf{(2) First Diffusion Model from Polarization to Other Modalities}: We design a diffusion model to achieve stable mapping from 16-channel Mueller matrix imaging to the stained image domain via a regulated diffusion process.

\textbf{(3) State-of-the-art Performance Validation}: Our model significantly outperforms other benchmark methods on our collected dataset in both polarization-to-H\&E and polarization-to-fluorescence tasks.
\section{Related Work}
\subsection{Image-to-image Translation}
Image-to-Image Translation (I2I) is an important research direction in the field of computer vision, aiming to transform images between different domains while maintaining the semantic consistency of the content. Recent advances in deep learning, particularly GANs and diffusion models, have significantly propelled progress in this field. Isola et al. proposed Pix2Pix, a Conditional-GAN (cGAN) framework for paired image translation, achieving success in tasks such as semantic segmentation and colorization~\cite{ref10}. Wang et al. extended this with Pix2PixHD, improving resolution and detail through a multi-scale architecture~\cite{ref11}. To realize structural fidelity, PyramidPix2pix~\cite{ref15} introduced a pyramid network architecture that progressively refines translations on multiple scales, effectively preserving hierarchical features in high-resolution outputs. 

Despite their success, GAN-based methods often suffer from training instability and mode collapse, limiting their ability to generate high-quality complex images. Recent advances in diffusion models have addressed these limitations. Li et al. introduced BBDM~\cite{ref17}, modeling I2I as a stochastic Brownian bridge process to directly learn domain transformations without iterative refinement. 
Xia et al. developed DiffI2I~\cite{ref19}, which introduces a compact I2I prior extraction network and dynamic I2I transformer to guide denoising, enabling fewer iterations and lighter models for high-quality translation. 


\subsection{Polarization Imaging in Virtual Staining}

Polarization imaging has emerged as a powerful tool in biomedical imaging due to its sensitivity to tissue micro-structural and anisotropic properties~\cite{ref20}. Recent advancements in polarization imaging have shown its potential in virtual staining~\cite{ref21}. Si et al.~\cite{ref22} proposed a deep learning-based approach using a cGAN structure to translate MM images into bright-field microscopy equivalents, addressing low throughput issues while preserving structural insights. Fan et al.~\cite{ref23} further simplified data acquisition by introducing an unsupervised CycleGAN framework trained on unpaired MM and H\&E-stained images, eliminating the need for spatial registration. 

These studies highlight the potential of polarization imaging in virtual staining, providing new auxiliary indicators for pathological diagnosis, and improving the value of multi-modality data through deep learning. 

\section{Multi-modality Polarization Pathology Dataset}
We present the Multi-modality Polarization Pathology Dataset (MPPD), the first breast cancer pathology dataset integrating Mueller matrix polarimetric imaging, H\&E staining, and fluorescence staining. The MPPD comprises 7 breast cancer patient samples, consisting of 18,000 pairs of 600×600 patches with precisely spatially aligned samples. This multi-modality framework enables the comprehensive characterization of pathological samples in structural, molecular, and physical domains.

\subsection{Data Collection}
To achieve high-efficiency Mueller matrix imaging, we developed an adaptive full-vector imaging system that integrates hardware control, image acquisition, and data analysis modules. The acquisition process begins with the use of specialized slide scanners to obtain Whole-Slide Images (WSIs) of H\&E staining and fluorescence staining. Subsequently, the samples are placed on a stage for polarimetric imaging. After autofocus determines the optimal focal plane, the system scans the sample region-by-region with a preset step size, recording the Mueller matrix data for each region. Given that samples are large and a single shot cannot cover the entire slice, an ITKMontage~\cite{ref28} module is employed for multi-modality image registration and stitching. To address issues such as size consistency and missing data in multi-modality image stitching, the system automatically generates configuration files to record tile position information. It also utilizes boundary detection and virtual tile generation techniques. Additionally, grayscale uniformity correction and brightness normalization are applied. Ultimately, this process results in the formation of complete panoramic WSI images under a 4$\times$ objective lens.

\begin{figure}[b] 
    \centering 
    \begin{minipage}[t]{0.3\textwidth}
        \centering
        Polarization
    \end{minipage}
    \hfill
    \begin{minipage}[t]{0.3\textwidth}
        \centering
        H\&E
    \end{minipage}
    \hfill
    \begin{minipage}[t]{0.3\textwidth}
        \centering
        Fluorescence
    \end{minipage}

    \begin{subfigure}[t]{0.3\textwidth} 
        \centering
        \includegraphics[width=\textwidth]{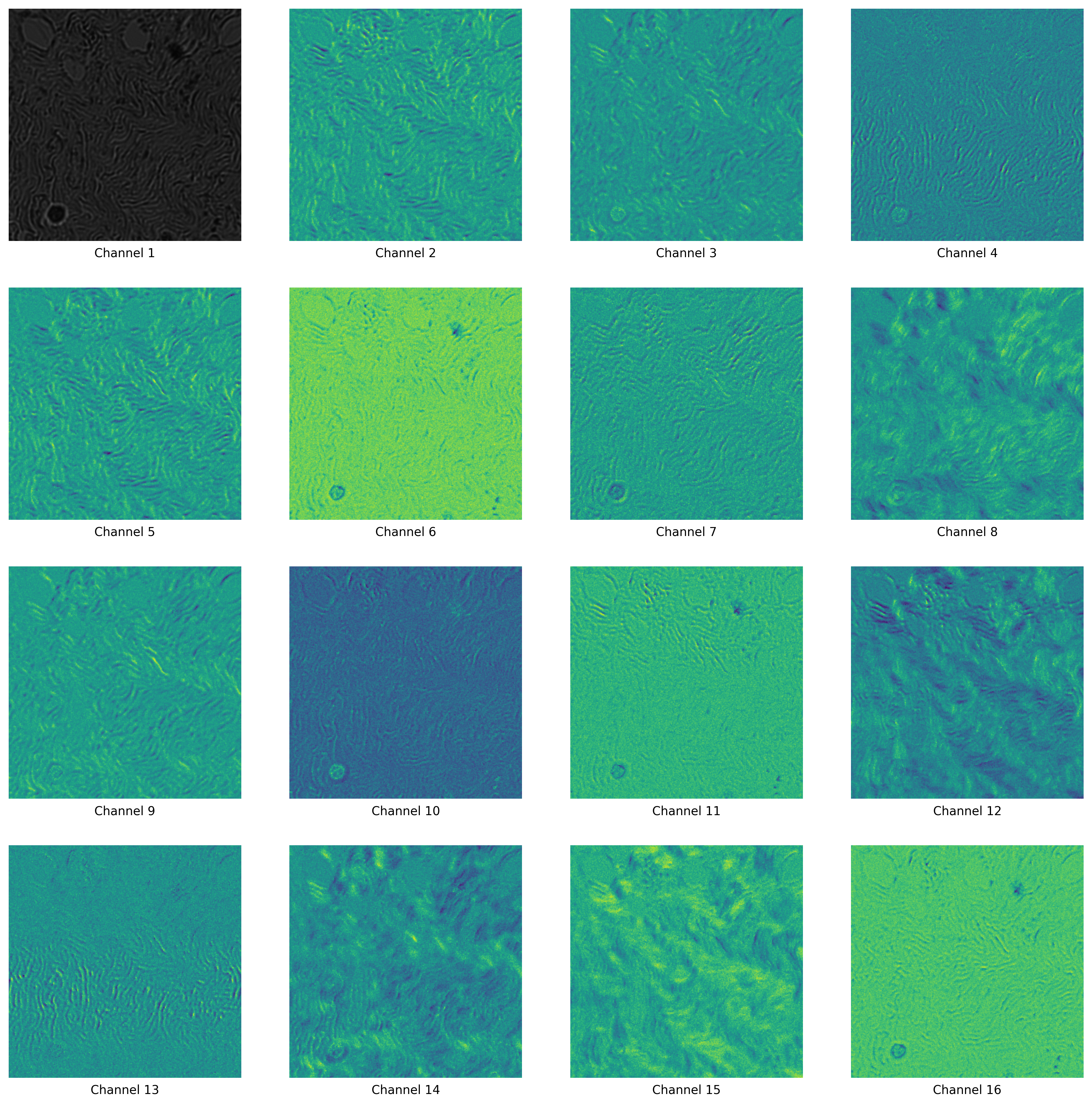}
        
    \end{subfigure}
    \hfill 
    \begin{subfigure}[t]{0.3\textwidth}
        \centering
        \includegraphics[width=\textwidth]{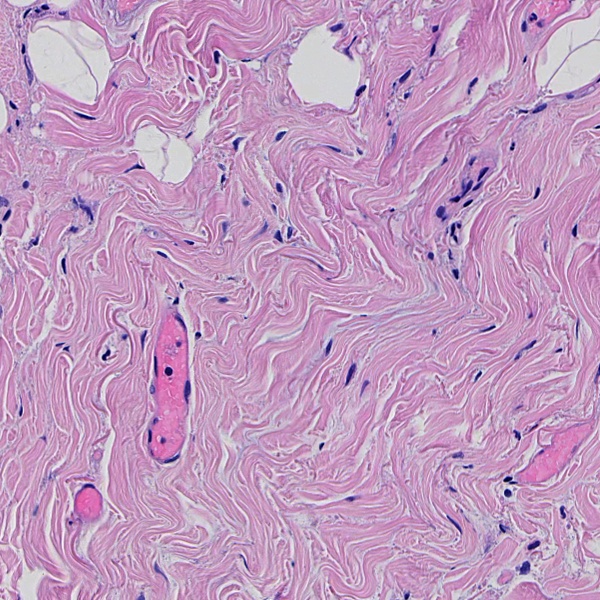}
    \end{subfigure}
    \hfill
    \begin{subfigure}[t]{0.3\textwidth}
        \centering
        \includegraphics[width=\textwidth]{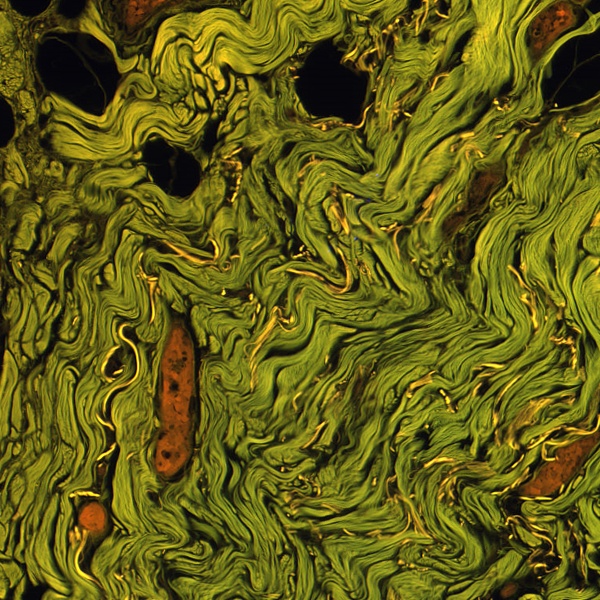}
    \end{subfigure}


    \caption{Polarization-H\&E-Fluorescence Dataset Visualization}
\end{figure}

\subsection{Data Processing} 
We designed a multi-stage registration pipeline to achieve precise spatial alignment for pixel-level correlation across modalities.Initial pre-processing standardizes the image scales and removes extraneous blank regions to mitigate feature interference. The polarization-derived gray modality is designated as the reference framework due to its consistent correspondence with other modalities. Feature matching is performed using the SuperPoint algorithm~\cite{ref29}, a deep learning-based detector that identifies cross-modal keypoints for robust correspondence estimation. Rigid registration optimizes global alignment via affine transformations. Subsequent non-rigid registration refines local deformations using B-spline modeling~\cite{ref30} combined with curvature regularization~\cite{ref31} to prevent artifacts. After registration, all modalities are resampled to 2304×1296 pixels and aligned with the polarization field-of-view (FOV). Then we applied adaptive thresholding to exclude tissue-free or artifact-dominated regions. The validated FOVs are partitioned into 600×600 pixel patches using a sliding window approach, preserving multi-modality alignment through strict spatial correspondence. This process produced 18,000 high-quality patches for downstream analysis.

\section{Methods}
We extend the Brownian Bridge Diffusion Models (BBDM)~\cite{ref17} to the polarization domain with additional regulations. Fig.~\ref{framework} shows the general structure of our Regulated Bridge Diffusion Model (RBDM).

\begin{center}
    \includegraphics[width=11.5cm, height=6cm]{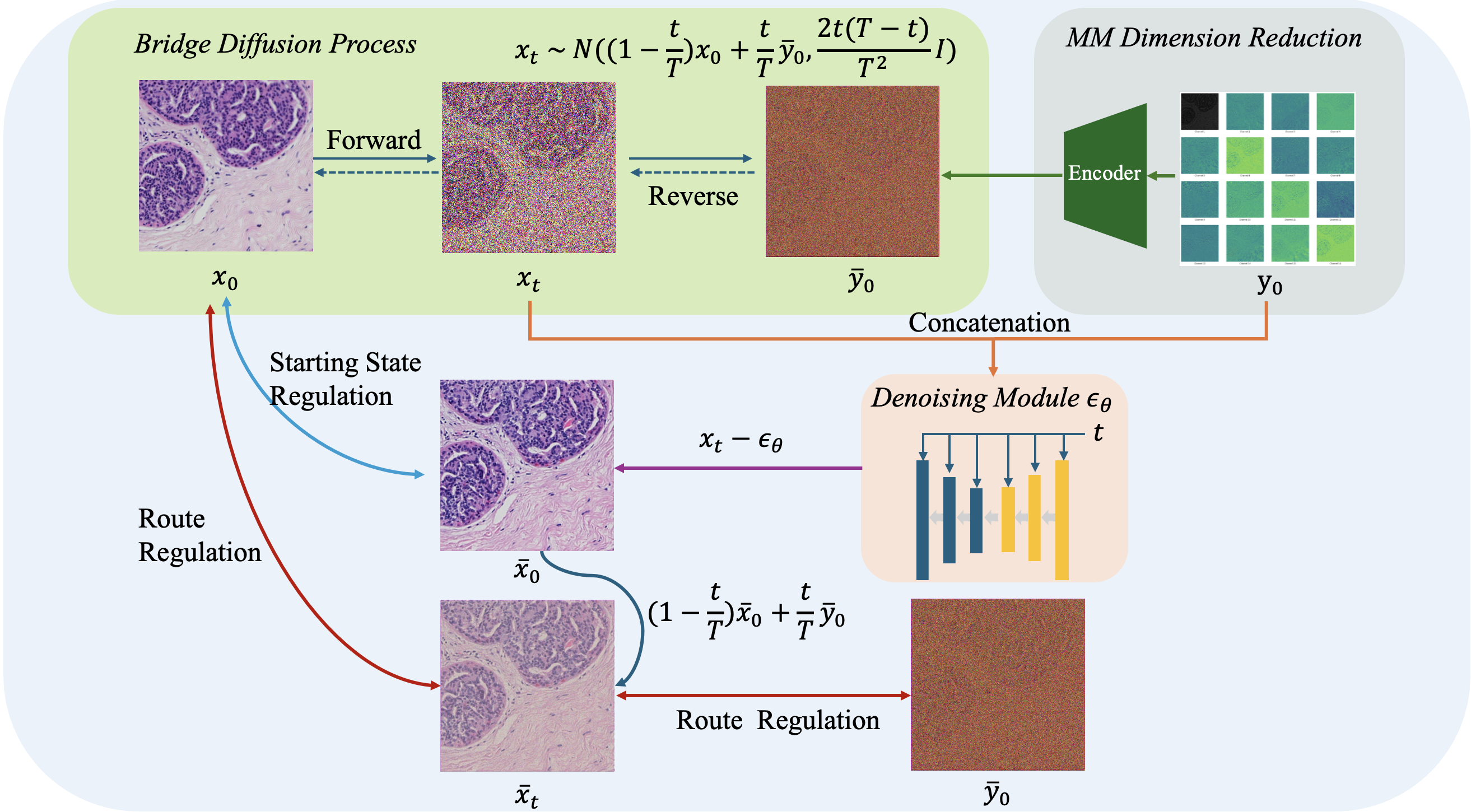} 
    \captionof{figure}{Overall Framework of Regulated Bridge Diffusion Model} 
    \label{framework} 
\end{center}

\subsection{Bridge Diffusion Model} 

Given an image $x_{0}$ from domain $A$ and image $y_{0}$ from domain $B$, the Bridge Diffusion Model aims to perform successful image-to-image translation from $y_{0}$ to $x_{0}$. During training, translation will go through the forward and reverse processes. In the forward process, the state distribution at each time step $t$ can be formulated as 
\begin{equation}
     q(x_{t}|x_{0}, y_{0})=N(x_{t};(1-\frac{t}{T})x_{0}+\frac{t}{T}y_{0}, \sigma _{t}I)
\end{equation}
Here $T$ is the total number of time steps, $\sigma _{t}$ stands for the variance, which is $\frac{2t(T-t)}{T^{2}}$, and it is easy to observe that at staring state and ending state, the $\sigma _{t}$ is $0$ and the mean value equals to $x_{0}$ and $y_{0}$ respectively, which makes it a fixed mapping process. And the intermediate state $x_{t}$ can therefore be represented as 
\begin{equation}
     x_{t} = (1-\frac{t}{T})x_{0}+\frac{t}{T}y_{0}+\sqrt{\sigma _{t}} \epsilon 
\end{equation}
where $\epsilon\sim N(0,I)$.
 
The reverse process starts with $y_{0}$ and aims to predict $x_{t-1}$ given $x_{t}$. It can be formulated as 
\begin{equation}
    p_{\theta } (x_{t-1}|x_{t}, y_{0})=N(x_{t-1};\mu_{\theta}(x_{t},t), \tilde{\sigma}_{t}I  )
\end{equation}
where $\tilde{\sigma}_{t}$ is the variance of the noise and $\mu_{\theta}(x_{t},t)$ is the predicted mean value of the noise. 

The training objective of the bridge diffusion model is to optimize the Evidence Lower Bound (ELBO) and is proven equivalent to training a network $\epsilon _{\theta }$ to predict noise, which can be formulated as 
\begin{equation}
   L _{BDM} = \left \| \frac{t}{T}(y_{0}-x_{0}) + \sqrt{\sigma_{t}}\epsilon - \epsilon_\theta(x_{t},t,y_{0})    \right \|  
\end{equation}
Once we know the $y_{0}$ and noise at each time step, $x_{0}$ can be reconstructed.

\subsection{Regulated Bridge Diffusion Model for Polarization} 
One main difference between polarization and other images is about the dimension, the traditional BBDM model cannot handle data with 16-dimensionality, so we design a simple auto-encoder structure $E_{p}$ for dimension reduction ra. The auto-encoder contains two convolutional layers, the first one expands the dimension from 16 to 64 and the second layer reduces it to 3, a tanh layer is appended to ensure the output scale is between -1 and 1, in this setting, the output of $E_{p}(y_{0})$ can be represented as $\bar{y}_{0}$, and the first training objective can be written as 
\begin{equation}
   L _{1} = \left \| \frac{t}{T}(\bar{y}_{0}-x_{0}) + \sqrt{\sigma_{t}}\epsilon - \epsilon_\theta(x_{t},t,y_{0})    \right \|  
\end{equation}
However, since the auto-encoder is trainable, during the training process, $\bar{y}_{0}$ may not be optimal. Therefore, we design two other regulation approaches for performance enhancement, denoted as starting state regulation and route regulation.

\subsubsection{Starting State Regulation}
Given any time step $t$, $x_{0}$ can actually be estimated by the following formula during training
\begin{equation}
\begin{split}
    \bar{x}_{0} & = x_{t} - \epsilon_{\theta}  (x_t,t,y_{0}) \\ 
    & = (1-\frac{t}{T})x_{0}+\frac{t}{T} \bar{y}_{0}+\sqrt{\sigma _{t}} \epsilon - \epsilon_{\theta}  (x_t,t,y_{0})
\end{split}
\end{equation} 
To ensure the estimated quality, we apply the perceptual loss~\cite{ref25} to narrow the difference between feature maps of $x_{0}$ and $\bar{x}_{0}$ using the pre-trained VGG~\cite{ref26} network. The starting state regulation loss can be expressed as 
\begin{equation}
    L _{2} = \left \| F(x_{0}) - F(\bar{x}_{0})\right \|  
\end{equation}
where $F$ stands for feature extraction network.

\subsubsection{Route Regulation}
The route regulation is according to the attribute of the Bridge Diffusion Model. When the time step $t$ is close to $T$, the intermediate image $\bar{x}_{t}$ generated should be more similar to $y_{0}$ than $x_{0}$, and when the time step $t$ is close to $0$, the image $\bar{x}_{t}$ should be more like $x_{0}$ without considering noise. MS-SSIM loss~\cite{ref27} is applied to measure the multi-scale structural similarity between the ground truth and the generated image, and the higher the value of the MS-SSIM is, the more similar the two images are. So the route regulation loss is then represented as 
\begin{equation}
    L_{3} = max(0, \frac{T-2t}{T}\left [  M(\bar{x}_{t},\bar{y}_{0})-M(\bar{x}_{t},x_{0}) \right ])
\end{equation}
Where $M$ means MS-SSIM value, and in order to train the whole network, we use $\bar{x}_{0}$ to generate new $\bar{x}_{t}$. The equation is 
\begin{equation}
    \bar{x}_{t}=(1-\frac{t}{T})\bar{x}_{0}+\frac{t}{T}\bar{y}_{0}
\end{equation}
Thus the overall training objective for our model becomes 
\begin{equation}
    L =  L _{1} + L _{2} + L _{3}
\end{equation}

\section{Experiments}
\subsection{Implementation Detail}
We use our original dataset for performance evaluation. 70\% patches are selected for training, and the rest are reserved for testing. All patches are resized to $256\times256$ and normalized to $\left [ {-1,1} \right ]$. For training, the total number of epochs is set to 40, and the Adam optimizer with a learning rate of $1e^{-4}$ is used. The batch size is 2 and the total number of time steps is set to 500 with a sample step of 100.
\subsection{Result}

\begin{figure}[b]
\centering
    \includegraphics[width=12cm, height=3.5cm]{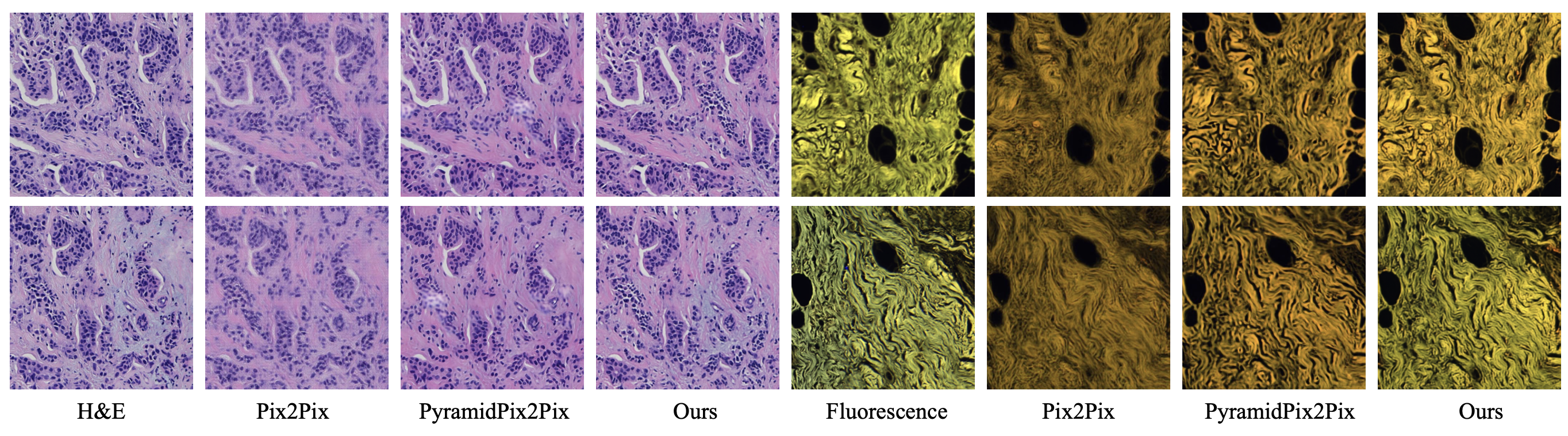} 
    \captionof{figure}{Visualization Results of Our Model and Other Baselines} 
    \label{results} 
\end{figure}

To make a fair comparison, we select multiple evaluation metrics and benchmark methods. The widely used metrics including Structural Similarity Index Measure (SSIM), Peak Signal-to-Noise Ratio (PSNR), Fréchet Inception Distance (FID) and Learned Perceptual Image Patch Similarity (LPIPS) are calculated to measure the image quality and similarity. The baseline methods are GAN-based model for paired data, including Pix2Pix~\cite{ref10} and PyramidPix2pix~\cite{ref15}. The results of polarization-to-H\&E and polarization-to-fluorescence are shown in Table~\ref{table1}. The task of transforming polarization to fluorescence is more difficult as the fluorescence is more complex. The best results are highlighted in bold. And it is obvious that our model significantly outperforms other methods in terms of all the evaluation indicators. We also picked four patches for the visualization of the image generation outcomes, the results are shown in Fig.~\ref{results}, the results from our model are more accurate and detailed compared to other baseline approaches.

\begin{table}[t]
\centering
\caption{Model Performance from Polarization to H\&E and Fluorescence} \label{table1}
\begin{tabular}{c|c|cccc}
Task                     & Model           & PSNR↑               & SSIM↑               & FID↓           & LPIPS↓        \\ \hline
                         & Pix2Pix         & 15.42±3.50          & 0.19±0.10          & 215.93         & 0.60          \\
Polar-\textgreater{}H\&E                 & Pyramid Pix2pix & 19.68±2.68          & 0.41±0.13          & 62.02          & 0.38          \\
                         & Ours            & \textbf{20.93±3.25} & \textbf{0.53±0.18} & \textbf{19.67} & \textbf{0.26} \\ \hline
                         & Pix2Pix         & 16.13±1.28          & 0.26±0.05          & 116.73         & 0.62          \\
Polar-\textgreater{}Fluo & Pyramid Pix2pix & 18.09±2.69          & 0.32±0.13          & 65.79          & 0.52          \\
                         & Ours            & \textbf{18.52±2.32} & \textbf{0.44±0.12} & \textbf{43.03} & \textbf{0.39}
\end{tabular}
\end{table}

\subsection{Ablation Study} 
To verify the effectiveness of the regulation modules, we also conduct the ablation study by adding the Starting State Regulation (SSR) and the Route Regulation (RR). The result is shown in Table \ref{table3}, which indicates that the diffusion-based model achieves better performance compared to previous baselines, and both SSR and RR help improve the performance of the model. 
\begin{table}[h]
\centering
\caption{Ablation Result of Starting State Regulation and Route Regulation} \label{table3}
\begin{tabular}{cc|cccc|cccc}
\multicolumn{2}{c|}{Module} & \multicolumn{4}{c|}{Polar-\textgreater{}H\&E}   & \multicolumn{4}{c}{Polar-\textgreater{}Fluo}    \\ \hline
SSR           & RR          & PSNR↑          & SSIM↑          & FID↓           & LPIPS↓        & PSNR↑          & SSIM↑          & FID↓           & LPIPS↓        \\ \hline
×             & ×           &  20.66±3.37              &   0.51±0.19             & 21.21               &  0.27             &    18.13±2.44            &  0.42±0.13              &  55.89              &  0.42             \\
\checkmark & ×           &   20.71±3.31             &    0.52±0.20            &   20.80             &  0.26             &  18.34±2.41              &          0.43±0.13      &   49.12             &    0.40           \\
\checkmark & \checkmark  & 20.93±3.25 & 0.53±0.18 & 19.67 & 0.26 & 18.52±2.32 & 0.44±0.12 & 43.03 & 0.39
\end{tabular}
\end{table}


\section{Conclusion}
In this paper, we extend the bridge diffusion model to the polarization domain and propose first public multi-modality polarization dataset named MPPD and a novel diffusion-based model named RBDM, which realizes the mapping from polarimetric Mueller matrix to standardized stained domain. We also conduct experiments on our MPPD datasets and the RBDM model achieves the state-of-the-art performance, which indicates a promising direction for polarization-based virtual staining in the future. 



%
%
%
%
\newpage
\bibliographystyle{splncs04}

\end{document}